\documentclass{nature}

\usepackage{graphicx}
\usepackage{epstopdf}
\DeclareGraphicsExtensions{.png,.pdf,.eps}
\usepackage{amsmath,amssymb}
\newcounter{firstbib}

\newcommand{\apj}{Astrophys. J.}
\newcommand{\apjl}{Astrophys. J.}

\newcommand{\aap}{Astron. Astrophys.}
\newcommand{\mnras}{Mon. Not. R. Astron. Soc.}

\newcommand{\nat}{Nature}

\newcommand{\pasp}{Publ. Astron. Soc. Pac.}

\title{A combined transmission spectrum of the Earth-sized exoplanets TRAPPIST-1\,b and c}

\author{Julien de Wit$^{1}$, Hannah R. Wakeford$^{2}$, Micha\"{e}l Gillon$^{3}$, Nikole K. Lewis$^{4}$, Jeff A. Valenti$^{4}$, Brice-Olivier Demory$^{5}$, Adam J. Burgasser$^{6}$, Laetitia Delrez$^{3}$, Emmanu\"{e}l Jehin$^{3}$, Susan M. Lederer$^{7}$, Amaury H. M. J. Triaud$^8$ \& Val\'{e}rie Van Grootel$^{3}$}

\begin{document}

\maketitle

\begin{affiliations}
 \item Department of Earth, Atmospheric and Planetary Sciences, Massachusetts Institute of Technology, 77 Massachusetts Avenue, Cambridge, Massachusetts 02139, USA;
 \item NASA Goddard Space Flight Center, Greenbelt, MD 20771, USA;
 \item Institut d'Astrophysique et de G\'{e}ophysique, Universit\'{e} de Li\`{e}ge, All\'{e}e du 6 Ao\^{u}t 19C, 4000 Li\`{e}ge, Belgium;
 \item Space Telescope Science Institute, 3700 San Martin Drive, Baltimore, MD 21218, USA;
 \item Astrophysics Group, Cavendish Laboratory, 19 J J Thomson Avenue, Cambridge, CB3 0HE, UK;
 \item Center for Astrophysics and Space Science, University of California San Diego, La Jolla, California 92093, USA;
 \item NASA Johnson Space Center, 2101 NASA Parkway, Houston, Texas, 77058, USA;
 \item Institute of Astronomy, Madingley Road, Cambridge CB3 0HA, UK.
\end{affiliations}
 

\newpage
\begin{abstract}
Three Earth-sized exoplanets were recently discovered close to the habitable zone\cite{Kopparapu2013,Zsom2013} of the nearby ultracool dwarf star TRAPPIST-1\cite{Gillon2016}. The nature of these planets has yet to be determined, since their masses remain unmeasured and no observational constraint is available for the planetary population surrounding ultracool dwarfs, of which the TRAPPIST-1 planets are the first transiting example. Theoretical predictions span the entire atmospheric range from depleted to extended hydrogen-dominated atmospheres\cite{Owen2013,Jin2014,Johnstone2015,Luger2015,Owen2016}. Here, we report a space-based measurement of the combined transmission spectrum of the two inner planets made possible by a favorable alignment resulting in their simultaneous transits on 04 May 2016. The lack of features in the combined spectrum rules out cloud-free hydrogen-dominated atmospheres for each planet at $\geq$10-$\sigma$ levels; TRAPPIST-1\,b and c are hence unlikely to harbor an extended gas envelope as they lie in a region of parameter space where high-altitude cloud/haze formation is not expected to be significant for hydrogen-dominated atmospheres\cite{morley2015}. Many denser atmospheres remain consistent with the featureless transmission spectrum---from a cloud-free water vapour atmosphere to a Venus-like atmosphere.
\end{abstract}



\begin{figure}
\begin{center}
\vspace{-2.5cm}\includegraphics[width=15cm,height=!]{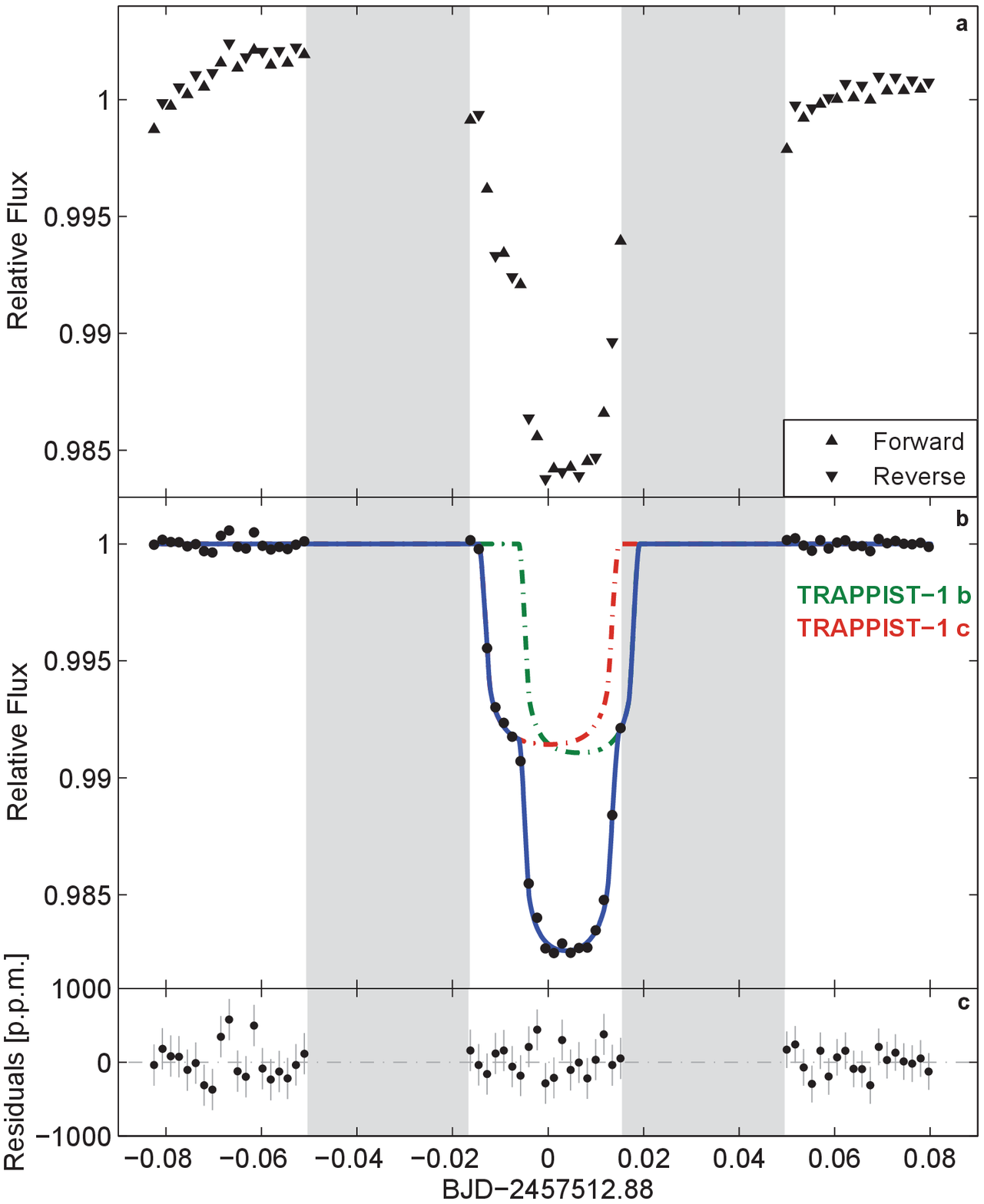}
\caption{Hubble/WFC3 white-light curve of TRAPPIST-1\,b and c double transit of 04 May 2016. {\bf a} Raw normalized white-light curve (triangles) highlighting the primary instrumental systematics (the forward/reverse flux offset and the ramp---see Methods). The shaded areas represent time windows during which no exposure was taken due to the Earth occultation. {\bf b} Normalized and systematics-corrected white-light curve (points) and best-fit transit model (line). The individual contribution of TRAPPIST-1\,b and c are shown in green and orange, respectively. {\bf c} Best-fit residuals with their 1-$\sigma$ error bars (SDNR = 240 p.p.m.).} \label{fig:1}
\end{center}
\end{figure}

\begin{figure}
\hspace{-1.5cm}\includegraphics[width=19cm,height=!]{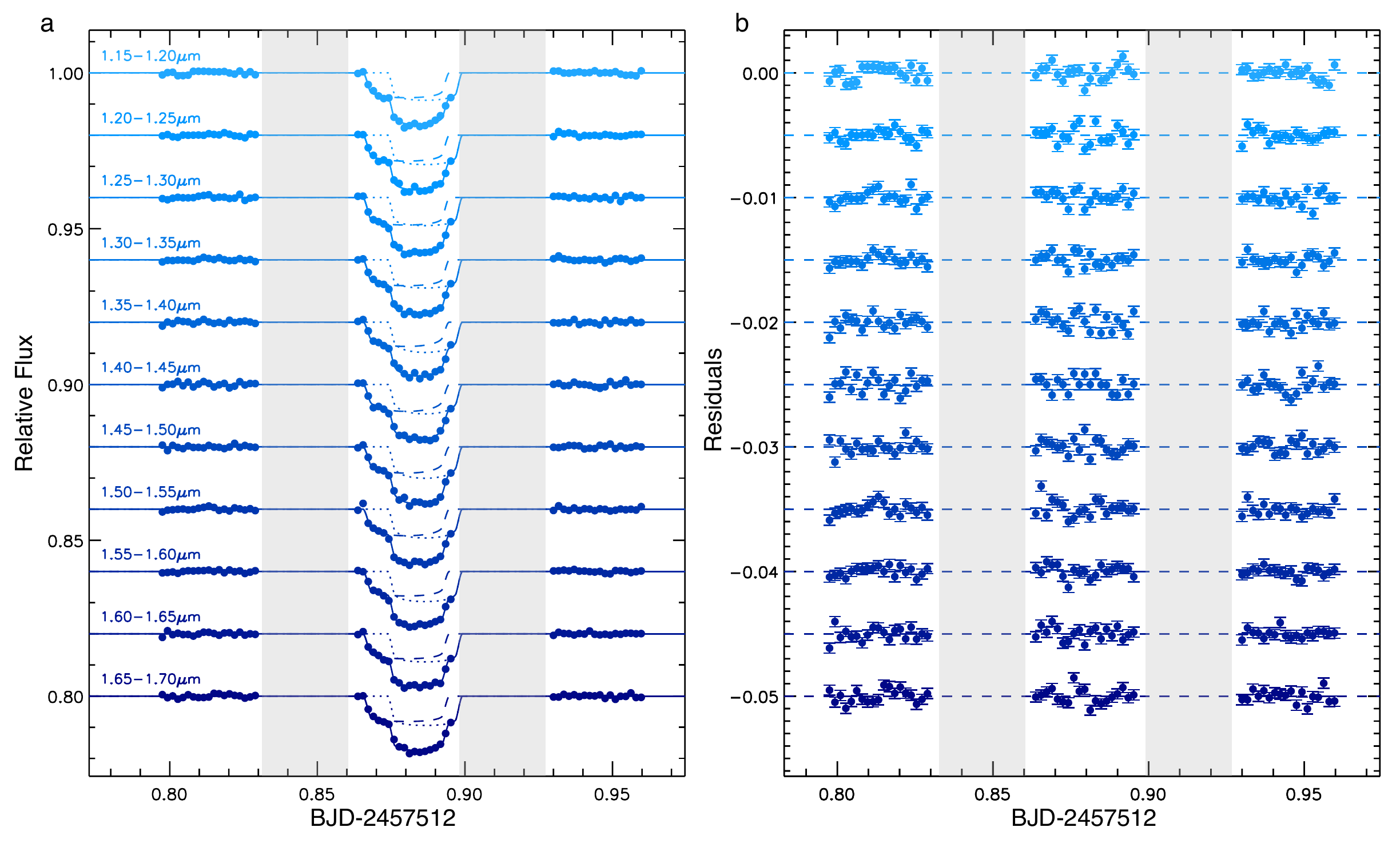}
\caption{Hubble/WFC3 spectrophotometry of TRAPPIST-1\,b and c double transit of 04 May 2016. {\bf a} Normalized and systematics-corrected data (points) and best-fit transit model (line) in 11 spectroscopic channels spread across WFC3 band, offset for clarity. The individual contribution of TRAPPIST-1\,b and c are shown in dotted and dashed lines, respectively. {\bf b} Best-fit residuals with their 1-$\sigma$ error bars (channel-averaged SDNR = 650 p.p.m.).} \label{fig:2}
\end{figure}

On May the fourth 2016, we observed the simultaneous transits of the Earth-sized planets TRAPPIST-1\,b and TRAPPIST-1\,c with the {\it Hubble Space Telescope} (HST). This rare event was phased with HST's visibility window of the TRAPPIST-1 system, allowing for the complete monitoring of the event (Figure\,\ref{fig:1}). Observations were conducted in ``Round-trip'' spatial scanning mode\cite{mccullough2012} using the near-infrared (1.1-1.7~$\mu$m) G141 grism on the Wide Field Camera 3 (WFC3) instrument (details in Methods). Following standard practice, we monitored the transit event through four HST orbits taking observations before, during, and after the transit event to acquire accurate stellar baseline flux levels. We discarded the first orbit due to differing systematics caused by the thermal settling of the telescope following target acquisition\cite{deming2013,wakeford2016,Sing2016}. The raw light curve presents primarily ramp-like systematics on the scale of HST orbit-induced instrumental settling discussed in previous WFC3 transit studies\cite{deming2013,Kreidberg2014,wakeford2016} (Figure\,\ref{fig:1}). We reduced, corrected for instrumental systematics, and analyzed the data using independent methods (see Methods) that yielded consistent results. We reached an average standard deviation of normalized residuals (SDNR) of 650 part per million (p.p.m.) per 112 s exposure (Figure\,\ref{fig:2}) on the spectro-photometric time-series split in 11 channels ($R = \lambda/\Delta\lambda \sim 35$). Summing over the entire WFC3 spectral range, we derived a ``white'' light curve with a 240 p.p.m. SDNR (Figure\,\ref{fig:1}).

We first analyzed the white light curve fitting for the transits of TRAPPIST-1\,b and TRAPPIST-1\,c simultaneously while accounting for instrumental systematics. Due to the limited phase coverage of HST observations, we fixed the system's parameters to the values provided in the discovery report\cite{Gillon2016} while estimating the transit times and depths. However, we let the band-integrated limb darkening coefficients (LDCs) and the orbital inclinations for planet b and c, $i_b$ and $i_c$ respectively, float under the control of priors, to propagate their uncertainties on the transit depth and time estimates with which they may be correlated. These priors were derived from the PHOENIX model intensity spectra\cite{Husser2013} for the LDCs (see Methods) and from the discovery report\cite{Gillon2016} for the planets' orbital inclinations. We find that TRAPPIST-1\,c initiated its transit 12 minutes before TRAPPIST-1\,b (transit time centers [BJD$_{TBD}$-2457512]: $T_{0,b}$ = 0.88646$\pm$0.00030 and $T_{0,c}$ = 0.88019$\pm$0.00016 ; transit durations\cite{Gillon2016} [min]: $W_b$ = 36.12$\pm$0.46 and $W_c$ = 41.78$\pm$0.81). The difference between the planets' transit duration of $5.6\pm0.9$ minutes implies that no planet-planet eclipse\cite{Hirano2012} occurred during the observed event, given the well-established orbital periods. Standard transit models\cite{Mandel2002} are therefore adequate for the analysis of the present dataset. We find an orbital inclination and transit depth across the full WFC3 band of $i_b = 89.39\pm0.32^\circ$ and  $\Delta F_b = 8015\pm220$ p.p.m. for TRAPPIST-1\,b and $i_c = 89.58\pm0.11^\circ$ and $\Delta F_c = 7290\pm240$ p.p.m. for TRAPPIST-1\,c. 

In the context of double transit observations, the data primarily constrain the combined transit depths ($\Delta F_{b+c} = 15320\pm160$ p.p.m.). Therefore, although the partial transit of TRAPPIST-1\,c---before TRAPPIST-1\,b initiates its transit---yields some constraints on $\Delta F_{c}$, it is not sufficient to completely lift  the degeneracy between $\Delta F_b$ ($=\Delta F_{b+c}-\Delta F_{c}$) and $\Delta F_c$. This explains the $\sim 30\%$ better precision obtained on the combined transit depth---and, hence, also on the combined transmission spectrum.  The transit depths derived over WFC3's band are in agreement within $\sim$2-$\sigma$ with the values reported at discovery\cite{Gillon2016}. 

\begin{figure}
\hspace{-2cm}\vspace{-2cm}\includegraphics[width=20cm,height=!]{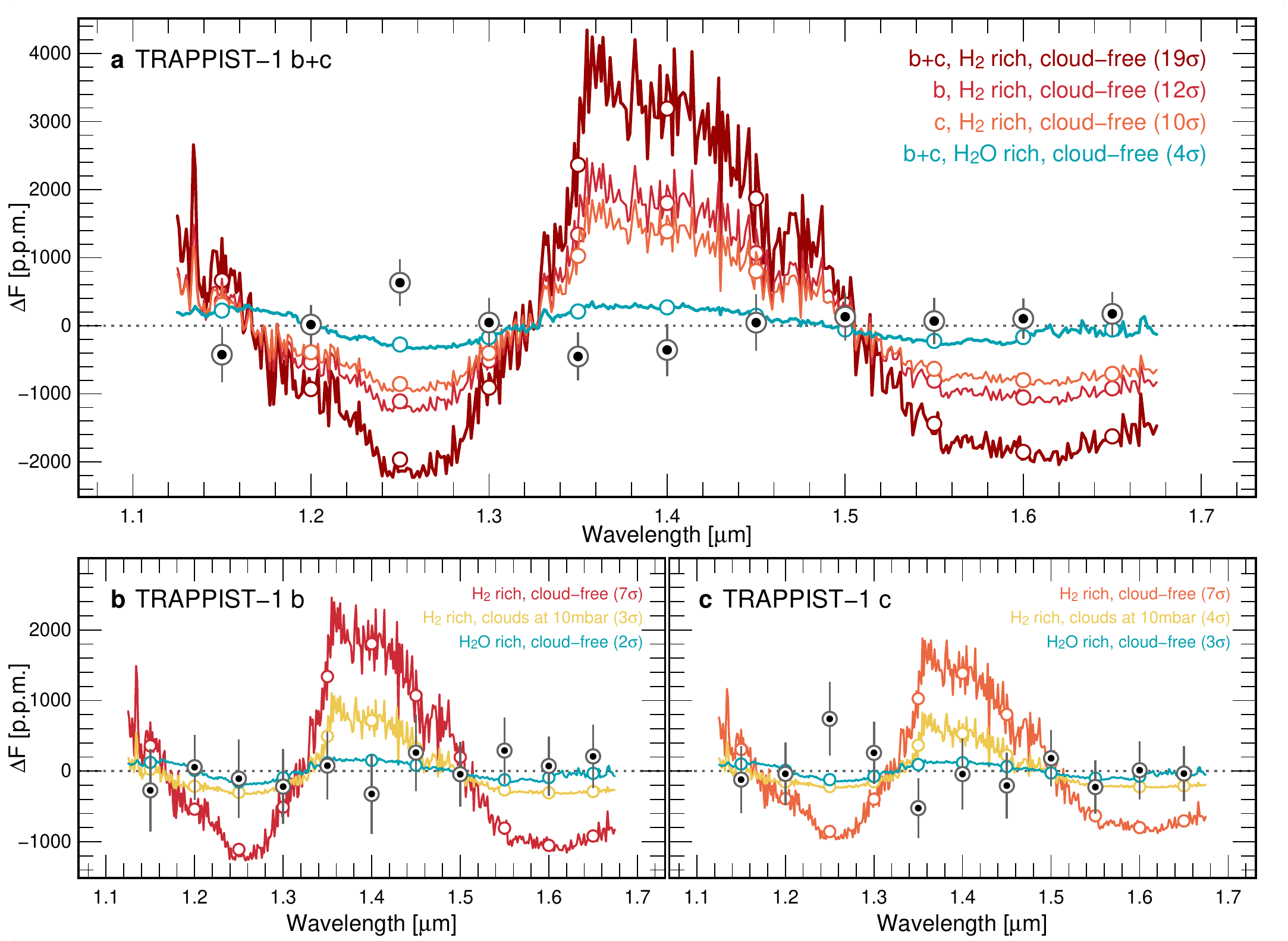}
\vspace{-0cm}\caption{Transmission spectra of TRAPPIST-1\,b and c compared to models. Theoretical predictions of TRAPPIST-1\,b's transmission spectrum {\bf b}, TRAPPIST-1\,c's {\bf c}, and their combinations {\bf a} are shown for cloud-free hydrogen-dominated (red lines and circles), hydrogen-dominated with a cloud deck at 10 mbars (yellow lines and circles), and cloud-free water-dominated (blue lines and circles) atmospheres. The colored circles show the binned theoretical models. The feature at 1.4$\mu$m arises from water absorption. The significance of the deviation of each transmission spectrum from WFC3 measurements (black circles with 1$\sigma$ errorbars) is listed in the legend in parentheses.}  \label{fig:3}
\end{figure}

We then analyzed the light curves in 11 spectroscopic channels, fitting for wavelength-dependent transit depths, instrumental systematics, and stellar baseline levels (Figure\,\ref{fig:2}). We tried both quadratic and 4-parameter limb darkening relationships\cite{sing2010} for each spectroscopic channel because transit depth estimates may depend on the functional form used to describe limb darkening. We found, however, that our conclusions are not sensitive to which limb darkening relationship was chosen, as long as the wavelength dependence of the LCDs is taken into account. The resulting transmission spectra are consistent with a flat line (Figure\,\ref{fig:3}). 

The transit depth variations expected over the WFC3 band if TRAPPIST-1\,b and/or TRAPPIST-1\,c were harboring a cloud-free hydrogen-dominated atmosphere are shown in Figure\,\ref{fig:3} (red lines and circles). Our transmission spectrum model\cite{deWit2013} sets atmospheric temperature to the planet's equilibrium temperature ($T_{eq,b} = 366$ K and $T_{eq,c} =315$ K) assuming a 0.3 Bond albedo. Since the planetary masses remain unmeasured, we conservatively use a mass of $\sim$0.95 and $\sim$0.85 M$_{\oplus}$ for TRAPPIST-1\,b and TRAPPIST-1\,c, the maximum masses allowing them to possess H$_2$-He envelopes greater than 0.1\% of their total masses given their radii \cite{howe2014}. The precision achieved on the combined transmission spectrum ($\sim$350 p.p.m. per bins) is sufficient to detect the presence of a cloud-free hydrogen-dominated atmosphere via the detection of water or methane absorption features. The featureless spectra rule out a cloud-free hydrogen-dominated atmosphere for TRAPPIST-1\,b and c at the 12 and 10$\sigma$ level, respectively. 

We also show in Figure\,\ref{fig:3} alternative atmospheres for TRAPPIST-1\,b and c that are consistent with the data; volatile (water) rich atmospheres and H$_2$-dominated atmospheres with a cloud deck at 10 mbar in blue and in yellow, respectively. Many alternatives for the atmospheres of TRAPPIST-1\,b and TRAPPIST-1\,c still remain. The atmospheric screening of sub-Neptune-sized exoplanets with current observatories is a step-by-step process\cite{Bean2010,Berta2012,Kreidberg2014}. As was done for the super-Earth GJ~1214\,b\cite{Bean2010}, the first observations of TRAPPIST-1's planets with HST allow us to rule out a cloud-free H$_2$-dominated atmosphere for either planet at the 10-$\sigma$ level. If the planets' atmospheres are hydrogen-dominated, then they must contain clouds or hazes that are grey absorbers between 1.1 and 1.7$\mu$m at pressures less than $\sim$10 mbar. However, theoretical investigations for hydrogen-dominated atmospheres\cite{morley2015} predict that the haze and cloud formation efficiencies at the irradiation levels of TRAPPIST-1\,b and TRAPPIST-1\,c should be dramatically reduced compared to, e.g., GJ~1214\,b (insolation ratios: $S_{GJ1214b}/S_{b} \sim 4$ and $S_{GJ1214b}/S_{c} \sim 8$), leading to cloud formation at pressure level $\geqq100$ mbar with marginal effects on their transmission spectrum\cite{deWit2013}. In short, H$_2$-dominated atmospheres can thus be considered as unlikely for TRAPPIST-1\,b and c.

Planets with the sizes and equilibrium temperatures of TRAPPIST-1\,b and c could possess relatively thick H$_2$O, CO$_2$, N$_2$, or O$_2$ dominated atmospheres or potentially tenuous atmospheres composed of a variety of chemical species \cite{Owen2013,Jin2014,Johnstone2015,leconte2015,Luger2015,Owen2016}. All these denser atmospheres are consistent with our measurements because the amplitude of a planet's transmission spectrum scales directly with its atmospheric mean molecular weight $\mu$---the amplitude of an exoplanet's transmission spectrum can be expressed as $2R_ph_{eff}/R^2_\star$, where $R_p$ and $R_\star$ are the planetary and stellar radii and $h_{eff}$ is the effective atmospheric height (that is, the extent of the atmospheric annulus) which is directly proportional
to the atmospheric scale height, $H = kT/\mu g$, where k is Boltzmann's constant, $T$ is the atmospheric temperature, and $g$ is the surface gravity. Therefore, everything else being equal the transmission spectrum amplitude of denser atmosphere is significantly damped compared to the one of a H$_2$-dominated atmosphere (e.g., by a factor $\sim$7 for a H$_2$O-dominated atmosphere). As a result, no constraint on the presence and minimum pressure-level of clouds/hazes for such denser atmospheres can be infered from our data. TRAPPIST-1\,b and c could, for instance, harbor a cloud-free water-vapor atmosphere or a Venus-like atmosphere with high-altitude hazes\cite{Tellmann2009,Wilquet2009}. We shall be able to distinguish between such atmospheres. The transmission spectrum of Venus as an exoplanet would present broad variations of $\sim$2 p.p.m. from 0.2 to 5 $\mu$m\cite{Ehrenreich2012} which, rescaled to the TRAPPIST-1 star, correspond to variations $\sim$160 p.p.m.($=2\times R^2_{Sun}/R^2_{TRAPPIST-1}$), currently below our errors, but eventually reachable.

Screening TRAPPIST-1's Earth-sized planets to progressively disentangle between their plausible atmospheric regimes and determine their amenability for detailed atmospheric studies will allow for the optimization of follow-up studies with the next generation of observatories. The present work highlights HST/WFC3's capability to perform the first step towards a thorough understanding of their atmospheric properties.

\newpage
\bibliographystyle{naturemag}


\begin{addendum}
 \item This work is based on observations made with the NASA/ESA Hubble Space Telescope that were obtained at the Space Telescope Science Institute, which is operated by the Association of Universities for Research in Astronomy, Inc. These observations are associated with program HST-GO-14500 (P.I.~J~de~Wit). H.R. Wakeford acknowledges support by an appointment to the NASA Postdoctoral Program at Goddard Space Flight Center, administered by USRA through a contract with NASA. M. G. is Research Associate at the Belgian F.R.S-FNRS science foundation. L. Delrez acknowledges support of the F.R.I.A. fund of the F.R.S-FNRS.  We thank D. Taylor, S. Deustua, P. McCullough, and N. Reid for their assistance in the planning and execution of our observations. We are also grateful for discussions with A. Burdanov, Z. Berta-Thompson, Pierre Magain, and Didier Queloz about the present study and the resulting manuscript. We thank the ATLAS and PHOENIX teams for providing stellar models.
 \item[Author Contributions] J.d.W and H.R.W. led the data reduction and analysis with the support of M.G., N.K.L., and B.-O.D. J.d.W., H.R.W., and N.K.L. led the data interpretation with the support of M.G. and J.A.V. J.A.V. provided the limb-darkening coefficients and further insights into TRAPPIST-1's properties and emission together with A.J.B. Every author contributed to write the manuscript and the HST proposal behind these observations.
 \item[Author Information] Reprints and permissions information is available at www.nature.com/reprints. The authors declare that they have no competing financial interests. Correspondence and requests for materials should be addressed to Julien de~Wit~(email: jdewit@mit.edu).
\end{addendum}

\newpage

\pagebreak





\newpage
\pagebreak
\clearpage
\begin{methods}

\subsection{HST WFC3 Observations.}
We observed the transit of TRAPPIST-1\,c followed 12 minutes later by the transit of TRAPPIST-1\,b on 04 May 2016. Observations were conducted using HST/WFC3 IR G141 grism (1.1-1.7$\mu$m) in Round-trip scanning mode\cite{mccullough2012}. Using the Round-trip scan mode involves exposing the telescope during an initial forward slew in the cross-dispersion direction, and exposing during an equivalent slew in the reverse direction (details on the tradeoffs behind round-trip scanning below). Scans were conducted at a rate of $\approx$0.236 pixels per second, with a final spatial scan covering $\approx$26.4 pixels in the cross-dispersion direction on the detector.

We use the IMA output files from the CalWF3 pipeline which have been calibrated using flat fields and bias subtraction. We applied two different extraction techniques which lead to the same conclusions. The first technique extracts the flux for TRAPPIST-1 from each exposure by taking the difference between successive non-destructive reads. A top-hat filter is then applied around the target spectrum measured $\pm$18 pixels from the center of the TRAPPIST-1 scan, and sets all external pixels to zero. Next, the images are reconstructed by adding the individual reads per exposure back together. Using the reconstructed images we extracted the spectra with an aperture of ±31 pixels around the computed centering profile for both forward and reverse scan observations. The centering profile is calculated based on the pixel flux boundaries of each exposure, which was found to be fully consistent across the spectrum for both scan directions. The second technique uses the final science image for each exposure and determines for each frame the centroid of the spectrum in a 28x136 pixel box, which corresponds to the dimensions of the irradiated region of WFC3's detector for the present observations. It then extracts the flux for 120 apertures of sizes ranging along the dispersion direction from 24 to 38 pixels (1 pixel increment) and along the cross-dispersion direction from 120 to 176 pixels (8 pixel increment)---we found the SDNR to be mostly insensitive to the aperture size along the dispersion direction. The best aperture was selected via a minimization of the SDNR of the white-light curve best fit, which is minimum for a 32x157 pixel aperture. Both techniques substract the background for each frame by selecting a region well away from the target spectrum, calculating the median flux, and cleaning cosmic ray detections with a customized procedure\cite{huitson2013}. Our observations present three cosmic ray detections that were not flagged by the CalWF3 pipeline. The exposure times were converted from Julian Date in universal time (JD$_{UT}$) to the Barycentric Julian Date in the Barycentric Dynamical Time (BJD$_{TDB}$) system \cite{Eastman2010}. Both extraction methods result in the same relative flux measurements from the star and SDNR ($\approx$ 240 p.p.m. in the white-light curve), as the build up of flux over successive reads is stable.

We elected to obtain the present observations using the Round-trip scan mode in order to increase the integration efficiency compared to the standard Forward scan mode. We note that due to slight differences in scan length/position and to the way the detector is read out (i.e., if the direction of the scan is in the same direction as the column readout then the integration time will be marginally longer than if the reverse was true\cite{mccullough2012}), Round-trip scan mode results in measurable differences in the total flux of the forward scan exposures compared to the reverse scan exposures. This effect has been seen for previous WFC3 observations\cite{knutson2014,Kreidberg2014} in Round-trip mode and has been corrected for in two main ways. The first method involves splitting the data into two sets, one for forward scan exposures and one for reverse scan exposures, effectively halving the number of exposures per light curve, but doubling the number of light curves obtained. Each of these datasets are then analysed separately and the results combined at the end\cite{Kreidberg2014}. The second method uses the median of each scan direction to normalize the two light curves which are then recombined and normalized prior to the light curve analysis to obtain the transit parameters\cite{knutson2014}. In the TRAPPIST-1 data we measure $\sim$0.1\% difference in flux level between the two scans. Due to the limited phase coverage of the combined transits, to retain the most information about the combined and separate effect of each planet, TRAPPIST-1\,c followed by the transit of TRAPPIST-1\,b, we cannot apply the first method. However, applying the second method we found significant remaining structure in the residuals suggesting that the correction is only partial. Previous observations using the Round-trip scan\cite{knutson2014} show that the offset between the lightcurve obtained with each scan varies significantly from orbit to orbit, suggesting that correcting via a median combine across visits is not optimal. In addition, the total flux is affected asymmetrically by other instrumental systematics---e.g., the detector ramp consistently yield a first measurement in the forward direction that is significantly lower than average---, biasing the median combine. Therefore, we corrected for the flux offset induced by the Round-trip scan mode based on the offset in the residuals for each HST orbit individually. To do so, we estimate in our forward model the ``intermediate residuals'' based on the data corrected for the transit model and the instrumental systematics, by the offset. For each orbit, we estimate the mean of the residuals for each scan direction ($m_f$ and $m_r$ for the mean of the residuals of the forward-scan exposures and the reverse-scan exposures, respectively). The ratio of the fluxes measured in reverse-scan exposures to the shared baseline level is $1+m_r$, and $1+m_r$ for forward-scan exposures. We therefore correct for their offsets by dividing each set of exposures by their respective ratio.

\subsection{HST WFC3 white-light curve and spectroscopy.}
We first analysed the white lightcurve by summing the flux across all wavelengths. We fitted the transits of TRAPPIST-1\,b and TRAPPIST-1\,c using the transit model of ref. \cite{Mandel2002} while correcting for instrumental systematics. We followed the standard procedure for the analyses of HST/WFC3 data by fixing the planets' orbital configurations---all but the orbital inclinations which are currently poorly constrained for TRAPPIST-1's planets---to the ones reported in the discovery report\cite{Gillon2016} while determining the transit times and depths. We used priors on the band-integrated limb darkening coefficients (LDCs) derived from the PHOENIX model intensity spectra\cite{Husser2013} and on the planets' orbital inclinations---these parameters being potentially correlated with the transit depth estimates---to adequately account for our current state of knowledge on TRAPPIST-1. We employed different analysis methods to confirm the robustness of our conclusions. The first method uses a least-squares minimization fitting (L-M) implementation \cite{wakeford2016} to investigate a large sample of systematic models which include corrections in time, HST orbital phase, and positional shifts in wavelength on the detector and marginalize over all possible combinations to obtain the transit parameters. It fits the lightcurves for each systematic model and approximates the evidence-based weight of each systematic model using the Akaike Information Criterion (AIC)\cite{Gibson2014}. It does so while keeping fixed the LDCs to the best estimates presented below and the orbital inclinations to the estimates from ref.\cite{Gillon2016}. The highest weighted systematic models include linear corrections in time, and linear corrections in HST orbital phase or in the shift in wavelength position over the course of the visit. Using marginalization across a grid of stochastic models allows us to therefore account for all tested combinations of systematics and obtain robust transit depths for both planets separately and in combination. For this dataset the evidence-based weight approximated for each of the systematic models applied to the data indicates that all of the systematic models fit equally well to the data and no one systematic model contributes to the majority of the corrections required to obtain the precision presented (Extended Data Figure 1). In other words, instrumental systematics affect marginally our observations. Independent analyses were carried out on the data using adaptive Markov Chain Monte Carlo (MCMC) implementations described in previous studies \cite{Gillon2012,deWit2016}.  For each HST light curve, the transit models\cite{Mandel2002} of TRAPPIST-1\,b and c are multiplied by baseline models accounting for the visit-long trend observed in WFC3 light curves, WFC3's ramp, and ``HST breathing'' effect\cite{wakeford2016}. For these analyses, priors are used for the LDCs and the orbital inclinations. The visit-long trend is found to be adequately accounted for with a linear function of time, the ramp with a single exponential in time, and the breathing with a second order polynomial in HST's orbital phase. More complex baseline models were tested and gave consistent results, as previously revealed by the marginalization study.

We calculated the transmission spectrum by fitting the transit depth of TRAPPIST-1\,b and TRAPPIST-1\,c simultaneously in each spectroscopic lightcurve. We divided the spectral range between 1.15-1.7$\mu$m into 11 equal bins of $\Delta\lambda$=0.05$\mu$m. We applied again the two techniques described above to analyse each spectroscopic light curve, resulting in the combined and independent transmission spectra of TRAPPIST-1\,b and TRAPPIST-1\,c. A least-squared minimization fitting (L-M) implementation \cite{wakeford2016} and the adaptive Markov Chain Monte Carlo (MCMC) implementations produced consistent results for each stage of the analysis.  

\subsection{Limb-darkening coefficients.}
We determined limb darkening coefficients by fitting theoretical specific intensity spectra ($I$) downloaded from the G\"ottingen Spectral Library\cite{URL_GSL}, which is described in ref.\cite{Husser2013}. The intensity spectra are provided on a wavelength grid with 1 \AA\ cadence for 78 $\mu$ values, where $\mu$ is the cosine of the angle between an outward radial vector and the direction towards the observer at a point on the stellar surface. We integrated $I$ over one broad and 11 narrow wavelength intervals used in our analysis of the transit light curve. We divided $I$ for each wavelength interval by $I_{\rm c}$, the value of $I$ at the center of the stellar disc ($\mu=1$). 

Because the PHOENIX code calculates specific intensity spectra in spherical geometry, the PHOENIX $\mu$ grid extends above the stellar limb relevant to exoplanet transit calculations. When fitting limb-darkening functions, PHOENIX $\mu$ values should be scaled to yield $\mu^\prime=0$ at the stellar radius\cite{espinoza2015}. We define $\mu^\prime=(\mu-\mu_0)/(1-\mu_0)$, where $I/I_{\rm c}=0.01$ at $\mu=\mu_0$. The value of $\mu_0$ is a function of wavelength. We then fitted two commonly used functional forms for limb-darkening\cite{sing2010},
$$ I/I_{\rm c} =
   1 - a(1-\mu^\prime)
     - b(1-\mu^\prime)^2
$$
and
$$
I/I_{\rm c} =
  1 - c_1(1-(\mu^\prime)^{1/2})
    - c_2(1-\mu^\prime)
    - c_3(1-(\mu^\prime)^{3/2})
    - c_4(1-\mu^\prime)^2.
$$
When fitting, we ignored points with $\mu^\prime < 0.05$.

Extended Data Figure 2 shows the limb darkening fits for the 12 wavelength intervals in our transit light curve analysis. We calculated fits for four stellar models with effective temperatures of 2500 and 2600 K and logarithmic surface gravities of 5.0 and 5.5. We then linearly interpolated the limb darkening coefficients to an effective temperature of 2550 and gravity 5.22, appropriate for TRAPPIST-1\cite{Gillon2016}.

\subsection{Transmission spectrum models.}
We simulated the theoretical spectra for TRAPPIST-1\,b and c using the model introduced in ref.\cite{deWit2013}. We used atmospheric temperatures equal to the planets' equilibrium temperature assuming a 0.3 Bond albedo (366K for TRAPPIST-1\,b and 315K for TRAPPIST-1\,c). The use of isothermal temperature profiles set at the equilibrium temperatures is conservative as it does not account for possible additional heat sources or temperature inversion and results in a possible under-evaluation of the atmospheric scale height. Our assumption on the temperature profiles does not impact our conclusion; variations of 50 K (i.e., $\sim$ 15\%) on the atmospheric temperature modify the amplitude of the transmission spectra by up to $\sim$ 15\% as at first order their amplitudes scale with the temperature. The planetary masses being unconstrained, we conservatively use a mass of $\sim$0.95 and $\sim$0.85 M$_{\oplus}$ for TRAPPIST-1b and TRAPPIST-1c, the maximum masses allowing them to possess H$_2$-He envelopes greater than 0.1\% of their total masses given their radii \cite{howe2014}. We used the atmospheric compositions of the ``mini-Neptune'' and ``Halley world'' models introduced in ref. \cite{benneke2012} to simulate the hydrogen-dominated and water-dominated atmospheres, respectively. We simulated the effect of optically thick cloud or haze at a given pressure level by setting to zero the transmittance of atmospheric layers with a higher pressure. 

The feature at 1.4$\mu$m arises from water absorption; the feature at 1.15$\mu$m for the water-dominated atmosphere arises from methane absorption. We compared the transmission spectra allowing for a vertical offset to account for our \textit{a priori} ignorance of the optically thick radius by setting the mean of each spectrum to zero. The significance of the deviation of each transmission spectrum from WFC3 measurements is listed in the legend in parentheses. Significance levels less than 3$\sigma$ mean that the data are consistent with that model within the reported errors. 

The presence of a cloud-free hydrogen-dominated atmosphere is ruled out for either planets at the 10$\sigma$ level by the combined transmission spectrum (and at a lesser 7$\sigma$ level by their individual spectra). The measurements are consistent with volatile (e.g., water) rich atmospheres or hydrogen-dominated atmospheres with optically thick clouds or hazes located at larger pressure than 10 mbar.

\subsection{Code availability.}
Conversion of the UT times for the photometric measurements to the BJD$_{TBD}$ system was performed using the online program created by J. Eastman and distributed at\\ 
http://astroutils.astronomy.ohiostate.edu/time/utc2bjd.html. We have opted not to make available the codes used for the data extraction as they are currently a significant asset of the researchers' tool kits. We have opted not to make available all but one of the codes used for the data analyses for the same reason. The MCMC software used by M.G. to analyse independently the photometric data is a custom Fortran 90 code that can be obtained upon request. The custom IDL code used to determine limb-darkening coefficients can be obtained upon request. 

\end{methods}

\newpage
\bibliographystyle{naturemag}

\newpage

\noindent Extended Data Figure 1: Marginal effects of the instrumental systematics on the transit depth estimates. {\bf a} Evidence-based weight, $W_q$ for each systematic model\cite{wakeford2016} applied to the white light curve. {\bf b} Combined transit depth estimate ($\Delta F_{b+c}$) obtained correcting the data using each systematic model. {\bf c} and {\bf d} show similarly the individual transit depth estimates of TRAPPIST-1\,b and c, $\Delta F_{b}$ and $\Delta F_{c}$ respectively. The horizontal lines indicate the final marginalized measurements and the associated uncertainties. The scale of the values here indicates that all of the systematic models fit equally well to the data.

\noindent Extended Data Figure 2: TRAPPIST-1's limb darkening. Stellar limb darkening relationships for TRAPPIST-1 (black curves) and four stellar models (colored curves) that bracket the effective temperature and surface gravity of TRAPPIST-1. The circles are theoretical \cite{Husser2013} specific intensities ($I$) relative to disc centre ($I_{\rm c}$) as a function of $\mu^\prime$, the cosine of the angle between an outward radial vector and the direction towards the observer. We fitted $I/I_{\rm c}$ averaged over the indicated wavelength intervals to determine the quadratic (dashed curves) and 4-parameter (solid curves) limb darkening coefficients.

\end{document}